\documentclass[11pt,a4paper]{article}
\usepackage[left=2.35cm,top=3cm,bottom=3cm,right=2.3cm]{geometry}
\usepackage{authblk}
\usepackage{amsmath}
\usepackage{amssymb}
\usepackage{cite}
\usepackage{tikz}
\usepackage{tabularx}
\usepackage{array}
\usepackage{diagbox}
\usetikzlibrary{patterns}
\usetikzlibrary{patterns.meta}
\usepackage{float}
\usepackage{comment}
\usepackage{}
\usepackage{subcaption}
\usepackage[nolist,nohyperlinks]{acronym}
\usepackage[colorlinks=true
,urlcolor=blue
,anchorcolor=blue
,citecolor=blue
,filecolor=blue
,linkcolor=blue
,menucolor=blue
,linktocpage=true
,pdfproducer=medialab
]{hyperref}
\usepackage{orcidlink}
\usepackage{slashed}

\def\dd{{\rm d}}

\usepackage{siunitx}
\usepackage{makecell}
\usepackage{booktabs}

\def\mcmule{{{\sc McMule}}}
\newcommand{\OL}{OpenLoops}
\newcommand{\GeV}{\mathrm{GeV}}
\newcommand{\MeV}{\mathrm{MeV}}
\newcommand{\rad}{\mathrm{rad}}

\newcommand{\vp}{\mathsf{p}}  

\catcode`,\active

\catcode`\,12

\definecolor{bluegray}{RGB}{140, 170, 250}
\definecolor{caribbeangreen}{rgb}{0.0, 0.8, 0.6}

\newcommand{\figref}[1]{Figure~\ref{#1}}
\newcommand{\tabref}[1]{Table~\ref{#1}}
\newcommand{\secref}[1]{Section~\ref{#1}}

\usetikzlibrary{decorations.markings,decorations.pathmorphing,positioning,decorations.pathreplacing,shapes,calc}
\tikzset{
    photon/.style={decorate, decoration={snake,amplitude=1pt,segment length=6pt}},
    zigzag/.style={decorate, decoration=zigzag},
}

\hypersetup{pdftitle={NNLO initial-state corrections to ee -> mu mu gamma},
pdfauthor={
Yizhou Fang,
Sara G\"undogdu,
Sophie Kollatzsch,
Marco Rocco,
Adrian Signer,
Yannick Ulrich}}

\title{
   NNLO initial-state corrections to $e^+\,e^-\to\mu^+\,\mu^-\,\gamma$  \\[5pt]}
\date{}
\author[1,2]{Yizhou~Fang\,\orcidlink{0009-0003-3568-9661}\,}
\author[1,2]{Sara~G\"undogdu\,\orcidlink{0009-0001-8780-7717}\,}
\author[1,2]{Sophie~Kollatzsch\,\orcidlink{0000-0002-8560-1619}\,\thanks{Present address: 
Max-Planck-Institut f\"ur Physik,
85748 Garching, Germany;
Physik Department, 
Technische Universit\"at M\"unchen, 
85748 Garching, Germany
}}
\author[3]{Marco~Rocco\,\orcidlink{0000-0002-2561-1209}\,}
\author[1,2]{Adrian~Signer\,\orcidlink{0000-0001-8488-7400}\,}
\author[4]{Yannick~Ulrich\,\orcidlink{0000-0002-9947-3064}\,}
\affil[1]{PSI Center for Neutron and Muon Sciences, 5232 Villigen PSI, Switzerland}
\affil[2]{Physik-Institut, Universit\"at Z\"urich, 8057 Z\"urich, Switzerland}
\affil[3]{Universit\`a degli Studi di Torino \& INFN, 10125 Torino, Italy}
\affil[4]{University of Liverpool, Liverpool L69 3BX, U.K.}

\allowdisplaybreaks

\begin{document}

\makeatletter
\begin{acronym}
    \newacro{LO}{leading order}
    \newacro{NLO}{next-to-leading order\def\AC@acronyms@NNLO{@<>@<>@}\def\AC@acronyms@LO{@<>@<>@}\def\AC@acronyms@NNNLO{@<>@<>@}}
    \newacro{NNLO}{next-to-next-to-leading order\def\AC@acronyms@NNNLO{@<>@<>@}\def\AC@acronyms@NLO{@<>@<>@}\def\AC@acronyms@LO{@<>@<>@}}
    \newacro{NNNLO}[N$^3$LO]{next-to-next-to-next-to-leading order\def\AC@acronyms@NLO{@<>@<>@}\def\AC@acronyms@LO{@<>@<>@}\def\AC@acronyms@NNLO{@<>@<>@}}
    \acro{PS}{parton shower}
    \acro{VFF}{vector form factor}
    \acro{VP}{vacuum polarisation}
    \acro{VPC}{\ac{VP} corrections}
    \acro{HVP}{hadronic vacuum polarisation}
    \acro{ISC}{initial-state corrections}
    \acro{FSC}{final-state corrections}
    \acro{qp}{quadruple precision}
\end{acronym}
\makeatother

\begin{titlepage}
\clearpage\maketitle 
\thispagestyle{empty}

\begin{abstract}\noindent
We present initial-state corrections to $e^+\,e^-\to\mu^+\,\mu^-\,\gamma$ at next-to-next-to-leading-order in QED for several experimentally motivated scenarios up to a few GeV.
We include photonic as well as vacuum-polarisation corrections into the Monte Carlo framework \mcmule{} that allows for the calculation of arbitrary infrared-safe observables. For the total cross sections, the corrections are typically below the percent level, but they can be locally larger in distributions. 
The vacuum-polarisation corrections are of similar size and must not be neglected, in particular for larger centre-of-mass energies.
\end{abstract}
\end{titlepage}

\section{Introduction} \label{sec:intro}
The ongoing RadioMonteCarLow~2 effort~\cite{Aliberti:2024fpq}
\begin{quote}
    \url{https://radiomontecarlow2.gitlab.io/}
\end{quote}
aims to improve predictions for $e^+ e^-$ colliders operating at centre-of-mass energies $\sqrt{s}$ of a few GeV, serving as input for several precision observables at low energies. One of the key processes is $e^+\,e^-\to \mu^+\,\mu^-\,\gamma$ which has played an important role for several experiments, such as Babar~\cite{BaBar:2012bdw}, KLOE~\cite{KLOE-2:2014qxg}, and BESIII~\cite{Ablikim:2020jrn}.
Ignited by this activity, predictions for this process, which in 2024~\cite{Aliberti:2024fpq} were available only at \ac{NLO}~\cite{Campanario:2013uea,Broggio:2022htr} or at \ac{LO} matched to a \ac{PS}~\cite{CarloniCalame:2003yt}, are pushed to the next order in perturbation theory.
Alongside the now available \ac{NLO}$+$\ac{PS} prediction~\cite{Budassi:2026lmr}, this work adds the first ingredient towards full \acs{NNLO} corrections to the most generic radiative QED process, $e^+e^- \to \mu^+\mu^-\gamma$.
Being a building block and a first technical step for many other processes at \acs{NNLO}, they will ultimately contribute to the \acs{NNNLO} corrections for $e^+e^-\to\mu^+\mu^-$, pushing the field of QED fixed order predictions to the next level.

Since the culmination of the first phase of the RadioMonteCarLow~2 effort~\cite{Aliberti:2024fpq}, there have been several advances in the description of $e^+e^- \to \mu^+\mu^-\gamma$ from Monte Carlo codes, in particular by \textsc{BabaYaga@NLO}~\cite{Budassi:2026lmr}, \textsc{Phokhara}~\cite{PetitRosas:2026ayp}, \textsc{Sherpa}~\cite{Price:2025fiu}.
They go along with efforts on the loop-integral and amplitude side~\cite{Badger:2022hno,Fadin:2023phc,Dave:2024ewl,Badger:2024fgb,PetitRosas:2025xhm,Pozzoli:2026eiu,Crisanti:2026luo}.
This work adds to the former and updates the fixed-order Monte Carlo framework \mcmule{}~\cite{Banerjee:2020rww} with \ac{ISC} for $e^+e^- \to \mu^+\mu^-\gamma$ together with the corresponding set of \ac{VPC}.
While \ac{ISC} only cover the gauge-invariant subset of radiative corrections to the initial-state particles of the process, they can be the dominant effects due to the collinear enhancement arising from 
the hierarchy in the lepton masses.
Moreover, \ac{ISC} considered as an off-shell process ($e^+e^-\to\gamma^*$) can be easily combined with any other, potentially hadronic, final state.

With this work, also the radiative process $2 \to 2 \,+ \,\gamma$ is pushed to the same perturbative accuracy as its $2\to 2$ counterpart.
In particular, for the process $e^+e^- \to \mu^+\mu^-$ (or its crossed version) the state-of-the-art is \acs{NNLO}~\cite{CarloniCalame:2020yoz,Budassi:2021twh,Broggio:2022htr,Aliberti:2024fpq} as well as all-order resummation-improved \ac{NLO}~\cite{Balossini:2008xr,Price:2025fiu,Price:2025pru}.

The results presented in this work are part of the publicly available version \texttt{v.0.8.0} of \mcmule{}
\begin{quote}
    \url{https://mule-tools.gitlab.io}
\end{quote}
Based on the all-order FKS$^\ell$ subtraction scheme~\cite{Engel:2019nfw} and the next-to-soft (NTS) stabilisation technique~\cite{Banerjee:2021mty}, \mcmule{} allows the calculation of any infrared-safe observable in a fully-differential way.
By exploiting the disperon QED method~\cite{Fang:2025mhn}, we are able to include leptonic and hadronic \ac{VP} insertions in a resummed form, including their contributions to loop diagrams.

This work is organised as follows.
In Section~\ref{sec:calc} we present technical details about the calculation.
In particular, we introduce our notation and define several subsets of the \ac{NLO} and \acs{NNLO} corrections.
This is followed by Section~\ref{sec:res}, where their size is shown for different scenarios ranging from $\sqrt{s}\simeq 1\,\GeV$ up to $10\,\GeV$.
Particular attention is paid to the quality of \ac{ISC} relative the full calculation and the impact of the \ac{VPC}.
We conclude in Section~\ref{sec:out} and comment on further work.

\section{Calculation} \label{sec:calc}
An overview of fixed-order calculations relevant to low-energy electron-positron scattering processes has been presented in~\cite{Aliberti:2024fpq}, whose notation and terminology we closely follow. As mentioned there, it is often convenient to organise contributions into gauge-invariant subsets according to their charge counting. To this end, we formally distinguish between the electron charge $q_e$ and the muon charge $q_m$. The \ac{LO} (differential) cross section for $e^+\,e^- \to \mu^+\,\mu^-\,\gamma$ therefore can be written as
\begin{align} \label{comp:loR}
\dd\sigma^{(0)}_{mm\gamma}(q^6)&=
\underbrace{\dd\sigma^{(0)}_{mm\gamma}(q_e^4\,q_m^2)}_{\text{ISC}}
+\underbrace{\dd\sigma^{(0)}_{mm\gamma}(q_e^2\,q_m^4)}_{\text{FSC}}
+\underbrace{\dd\sigma^{(0)}_{mm\gamma}(q_e^3\,q_m^3)}_{\text{mixed}}\,  .
\end{align}
In the \ac{ISC} (\ac{FSC}) part, the photon is emitted from the electron (muon) line, whereas the mixed term is due to the interference. The \ac{NLO} corrections $\dd\sigma^{(1)}_{mm\gamma}(q_e^i\,q_m^j)$ contain two additional powers of $q$, with $i+j=8$ and $2\le i,j \le 6$. The \ac{NLO} \ac{ISC} are $\dd\sigma^{(1)}_{mm\gamma}(q_e^6\,q_m^2)$. In addition, there are \ac{VP} insertions $\dd\sigma^{(1)}_{mm\gamma}(q_e^i\,q_m^j\,\Pi)$ with $i+j=6$ and $2\le i,j \le 4$. From a strict QED point of view, at \ac{NLO} the one-loop (leptonic) \ac{VP} $\Pi_\ell^{(1)}$ needs to be inserted. From a more phenomenological point of view, it is crucial to also insert the \ac{HVP} $\Pi_h$ and often resum the sum of them.  
All these contributions have been computed by various collaborations and a technical comparison and validation was done in~\cite{Aliberti:2024fpq}.

\begin{figure}[t]
    \centering
    \includegraphics[width=0.95\textwidth]{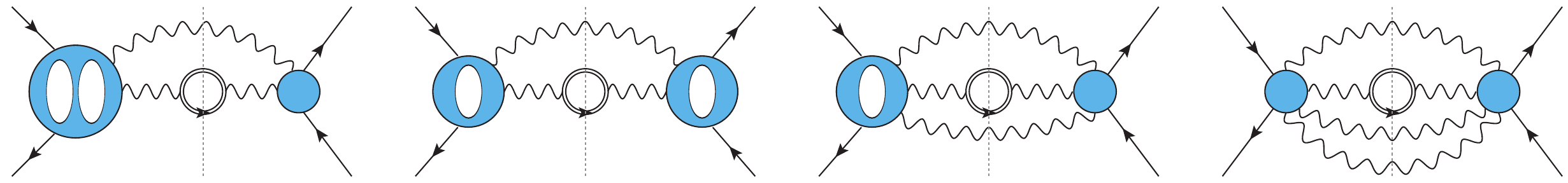}
    \caption{Schematic photonic \ac{NNLO} \ac{ISC} to $e^+\,e^- \to \mu^+\,\mu^-\,\gamma$, from left to right: Interference of two-loop and tree-level amplitude, one-loop amplitude squared, real-virtual contribution, real-real contribution. 
    The number of holes in the blue vertices represents the loop order.
    }
    \label{fig:diags}
\end{figure}

Here we present \ac{NNLO} \ac{ISC} to $e^+\,e^- \to \mu^+\,\mu^-\,\gamma$, i.e.~all contributions with the counting $\dd\sigma^{(2)}_{mm\gamma}(q_e^8\,q_m^2)$. We call them the photonic \ac{ISC}. As illustrated in \figref{fig:diags}, they have virtual-virtual, real-virtual, and real-real contributions that will be discussed further in \secref{sec:calc-ph}. They can be seen as photonic \ac{NNLO} corrections to the process $e^+\,e^-\to\gamma\,\gamma^*$ with a subsequent \ac{LO} `decay' $\gamma^*\to\mu^+\,\mu^-$.

Following this line of thought, in \secref{sec:calc-vp} we also present \ac{NNLO} \ac{VPC} to $e^+\,e^-\to\gamma\,\gamma^*$. There are two classes of diagrams to be considered, which we dub non-factorising and factorising \ac{VP} insertions respectively. The first -- illustrated in the left panel of \figref{fig:vp} -- can be visualised by considering the one-loop correction to this process, with an additional \ac{VP} insertion for the photon in the loop.
As depicted in the right panel of \figref{fig:vp}, there are also contributions where the \ac{VP} is inserted in the $\gamma^*$ line connecting the electron and muon. These factorising insertions are trivial to compute as the loop integration is not affected. As mentioned above, they are often already included in the \ac{NLO} result through resummed \ac{VPC}. To be consistent at \ac{NNLO}, the resummation has to include the two-loop leptonic \ac{VP} $\Pi_\ell^{(2)}$ as well. 

For all these \ac{VPC}, there are only virtual-virtual contributions.
We do not consider the additional production of a lepton pair because the process $e^+\,e^-\to\mu^+\,\mu^-\,\gamma\, \ell^+\,\ell^-$ can in principle be distinguished. Thus, it might be a reducible background to a particular observable but is clearly separate from the process we consider. We also do not consider the light-by-light contributions due to a fermion (hadron) box in the two-loop amplitude of $e^+\,e^-\to\gamma\,\gamma^*$. These contributions were studied for the process $e^+\,e^-\to\gamma\,\gamma$ in~\cite{Engel:2025fmp} and shown to be negligible compared to other \ac{NNLO} corrections. 

\subsection{Photonic corrections} \label{sec:calc-ph}

The computation of the photonic \ac{NNLO} \ac{ISC} follows the standard procedure used by \mcmule{} for several $2\to{2}$ \ac{NNLO} results, as described in~\cite{Broggio:2022htr}. 
All parts illustrated in \figref{fig:diags} contribute. 
The ultraviolet singularities are eliminated through renormalisation and all results are expressed in terms of the on-shell lepton masses and on-shell coupling $\alpha$. The soft infrared singularities -- regulated through dimensional regularisation -- are isolated and cancelled between the virtual and real corrections using the FKS$^\ell$~\cite{Engel:2019nfw} subtraction method. 

In principle, the full dependence on the electron mass is kept. Hence collinear emission does not produce singularities but leads to a logarithmic enhancement that forms a crucial part of the results. There is one part in the calculation, where the full $m_e$ dependence is not available. This concerns the virtual-virtual correction.
In particular, the two-loop amplitudes (to be interfered with the tree-level amplitude, as indicated in the first diagram of \figref{fig:diags}) is only known with vanishing electron mass $m_e$~\cite{Garland:2001tf,Garland:2002ak,Badger:2023xtl,Fadin:2023phc}. 
We use the expressions of~\cite{Badger:2023xtl} and apply massification~\cite{Penin:2005eh,Becher:2007cu,Engel:2018fsb} to obtain all terms that are not polynomially suppressed by $m_e/Q$, where $Q$ is any other typical scale of the process. This leads to an error of the order $(\alpha/\pi)^2\,m_e^2/Q^2$, possibly multiplied by a logarithm $\log(m_e^2/Q^2)$. 
In the scenarios we will consider in \secref{sec:res}, there is always a tagged photon emitted at a large angle w.r.t.~the beam. Hence, the typical $Q \simeq\sqrt{s}$ and the numerical impact of the massification error is completely negligible. However, for observables with no tagged photon, the latter can be emitted with a momentum $k$ collinear to an incoming electron $p_i$. In this case $Q^2\equiv (p_i+k)^2 \sim m_e^2$ is possible and the massification error needs to be assessed more carefully. Hence, in this work we will not consider such scenarios.

The real-virtual corrections require the one-loop matrix elements with two photons. They are evaluated with full mass dependence using \OL~\cite{Buccioni:2017yxi, Buccioni:2019sur}. These matrix elements have to be integrated over the phase space of the additional photon, including the region where it is soft and collinear to the electrons. In order to improve the stability of the integration in an efficient way, we use NTS stabilisation~\cite{Banerjee:2021mty}. This is achieved by using the one-loop extension~\cite{Engel:2021ccn} of the LBK theorem~\cite{Low:1958sn, Burnett:1967km} to approximate the matrix element with a soft photon at next-to-leading power in the soft photon-energy expansion. 

Finally, the real-real corrections require the matrix elements with two additional photons, to be integrated over their phase space. This part is also done with full mass dependence, using \OL{} or an in-house calculation for the matrix elements.

As described in~\cite{Engel:2019nfw}, the individual parts of the \ac{NNLO} cross section depend on an unphysical parameter $\xi_c$. The cancellation of the $\xi_c$ dependence between real and virtual corrections is a useful check on the implementation of the subtraction method and on the numerical stability of the code.

\subsection{Fermionic or vacuum-polarisation corrections} \label{sec:calc-vp}

\begin{figure}[t]
    \centering
    \includegraphics[width=0.5\textwidth]{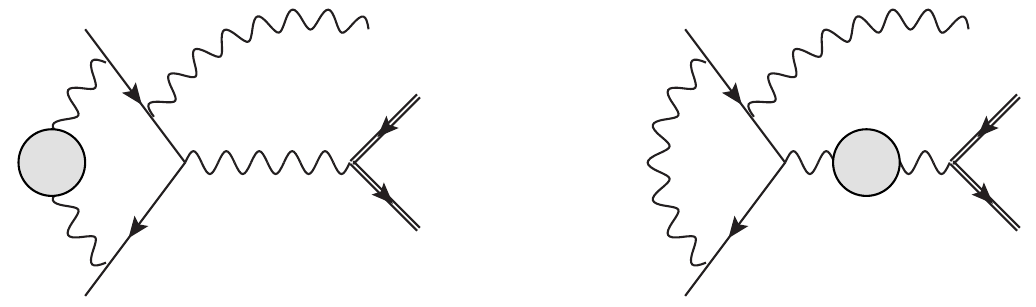}
    \caption{Illustrative examples of \ac{VPC} diagrams for $e^+\,e^- \to \mu^+\,\mu^-\,\gamma$. The left panel shows an example of a non-factorising \ac{VP} insertion in a loop propagator, whereas the right panel shows a factorising VP insertion in a photon that is not part of the loop. }
    \label{fig:vp}
\end{figure}

From a pure QED point of view, fermionic corrections are those containing fermion loops. This notion is typically kept, even if hadronic loops are included. Since we neglect fermionic box contributions (i.e.~light-by-light), the fermionic corrections correspond to \ac{VP} insertions.

For \ac{ISC}, the non-factorising \ac{VPC} which are genuinely \ac{NNLO} consist of two-loop diagrams for $e^+\,e^-\to\gamma\,\gamma^*$ that contain a \ac{VP} inserted into a photon inside the loop, see left panel of \figref{fig:vp}.
For leptonic \ac{VP} with massless fermions the results are typically included in the two-loop matrix elements for $e^+\,e^-\to\mu^+\,\mu^-\,\gamma$. However, these results are not sufficient for our purpose. First, we want to keep non-vanishing fermion masses wherever possible. Second, and more important, the \ac{HVP} is a complicated non-perturbative function that is inferred from experimental data.

In order to include a generic \ac{VP}, we use disperon QED~\cite{Fang:2025mhn} and express the loop integrals involving the \ac{VP} through a dispersion relation in terms of canonical loop integrals. This entails an additional integration over a dispersive parameter that is carried out together with the phase-space integration. This is implemented through the adapted version of \OL{} presented in~\cite{Fang:2025mhn}. Our set up allows to consider separately the leptonic \ac{VP} (dominated by electron loops) and the (strongly suppressed) \ac{HVP} as well as their resummed version. 
For the \ac{HVP}, any external function can be used. Our default choice for the results in \secref{sec:res} is~\cite{Ignatov:hvp}.

There are also \ac{VPC} which factorise as they appear in the $\gamma^*$ line that connects to the muon, see right panel of \figref{fig:vp}.
These include the product of a single \ac{VP} insertion for the one-loop leptonic $\Pi_\ell^{(1)}$ and a hadronic \ac{VP} $\Pi_h$, as well as the genuine
two-loop self-energy for the leptons $\Pi_\ell^{(2)}$.
The analytic result for the latter is obtained from~\cite{Djouadi:1993ss}.

For both types of \ac{VPC} as well as at \ac{NLO}, we will include them resummed. To this end we use the standard modification of the photon propagator, including the geometric series of $\Pi_2\equiv \Pi_\ell^{(1)}+\Pi_\ell^{(2)}+\Pi_h$ insertions. This is trivial to do for factorising corrections. 
Using disperon QED, it is in fact also straightforward to do for non-factorising \ac{VPC}. The only modification required is to adapt the dispersive representation of the modified photon propagator. 

\section{Results} \label{sec:res}

With the calculation described in the previous section implemented in \mcmule{} we are able to compute any infrared-safe observable in a fully differential way. 
In this section we present some observables that have already been considered in the RadioMonteCarLow~2 effort~\cite{Aliberti:2024fpq}. 

Following the notation of~\cite{Aliberti:2024fpq}, the momenta, polar and azimuthal angles of the muons are denoted by $p_\pm=(E_\pm, \vec{p}_\pm)$, $\theta^\pm$ and $\phi^\pm$. Further, we use $\vp_\pm\equiv|\vec{p}_\pm|$ while $\vp^z_\pm$  and $\vp^\perp_\pm$ denote the magnitude of the $z$ and transverse components of the charged final-state particles. An analogous notation is used for photons. In the experimental scenarios we consider here, there is always at least one tagged photon. However, beyond \ac{LO} more than one photon might be in the final state. In this case we assume they can always be separated.  The hardest photon passing the cut is denoted by $\gamma^{(h)}$. We also define $\vec{p}_{\widetilde\gamma}\equiv -(\vec{p}_{+}+\vec{p}_{-})$. This corresponds to the total 3-momentum of all emitted photons. In addition, we make use of the following quantities:
\begin{subequations} 
\begin{align}
   &\mbox{invariant mass}&  &M_{\mu\mu}\equiv \sqrt{(p_++p_-)^2} \quad \mbox{or} \quad 
      M_{\mu\mu\gamma}\equiv \sqrt{(p_++p_-+p_{\gamma^{(h)}})^2}  \\
    &\mbox{(polar angle) acollinearity}& &\xi\equiv|\theta^+ + \theta^- - \pi|& \\
     &\mbox{average polar angle}  & & \theta_\text{avg}\equiv (\theta^- - \theta^+ + \pi)/2 & \\
    &\mbox{azimuthal angle acollinearity} &  &\big| |\phi^+ - \phi^-| - \pi\big|& \\
     &\mbox{track mass} &     &M^2_{\rm trk}=\frac{a}{4}-\frac{\vp_-^2+\vp_+^2}{2}+\frac{(\vp_+^2-\vp_+^2)^2}{4a}& \label{def-mtrk}
\end{align}
\end{subequations}
The track mass in~\eqref{def-mtrk} uses the abbreviation $a=(\sqrt{s}-|\vec{p}_-+\vec{p}_+|)^2$ and is defined as the solution of
\begin{align}
    \left(\sqrt{s}-\sqrt{\vp_+^2+M_{\rm trk}^2}-\sqrt{\vp_-^2+M_{\rm trk}^2}\right)^2-\left(\vec{p}_++\vec{p}_-\right)^2\equiv 0 \,.
\end{align}
For observables at $\text{N}^{n}\text{LO}$, we introduce the notation
\begin{align}
    \sigma_n=\sigma^{(0)}\,+\,\sigma^{(1)}\,+\,...\,+\sigma^{(n)}\,.
\end{align}
Further, we define the $K$ factors that show the impact of higher-order \ac{ISC} ($K_{\rm ISC}^{(i)}$) or full corrections ($K^{(i)}$)
\begin{align}
    \label{eq:Kfactor}
    K^{(i)} = 1 + \delta K^{(i)} = \frac{\sigma_{i}}{\sigma_{i-1}}, \quad \text{and} \quad K_{\rm ISC,x}^{(i)} = 1 + \delta K_{\rm ISC,x}^{(i)} = \frac{\sigma_{i-1} + \sigma^{(i)}_{\rm ISC,x}}{\sigma_{i-1}},
\end{align}
with the optional $x\in\{\text{VPC}, \text{photonic}\}$. As discussed in \secref{sec:calc-vp}, the full \ac{NLO} correction includes the resummation of 
$\Pi_2$
in the intermediate photon propagator. 
In order to further decompose and show the impact of the individual \ac{NNLO} \ac{VPC}, we define
\begin{align}
    \label{eq:defVPdeltas}
    \delta^{(2)}_{\rm VPC} = \frac{
    \sigma^{(2)}_{\rm ISC, VPC}
    }{
    \sigma^{(2)}_{\rm ISC, photonic}
    } =
    \frac{
    \sigma^{(2)}_{\rm ISC, VPC}\big\vert_{\rm fac}
    }{
    \sigma^{(2)}_{\rm ISC, photonic}
    } +
    \frac{
    \sigma^{(2)}_{\rm ISC, VPC}\big\vert_{\text{non-fac}}
    }{
    \sigma^{(2)}_{\rm ISC, photonic}
    } 
    = \delta^{(2)}_{\rm VPC}[{\rm fac}] + \delta^{(2)}_{\rm VPC}[\text{non-fac}]\,.
\end{align}
Again, we always resum $\Pi_2$ insertions in the corresponding photon propagators.

All raw data and analysis pipelines can be found at~\cite{McMule:data}
\begin{quote}
    \url{https://mule-tools.gitlab.io/user-library/dilepton/}
\end{quote}
in the corresponding folders named after the scenarios defined later.

Before presenting differential distributions for the invariant mass of the muon pair $M_{\mu\mu}$ and the average polar angle $\theta_{\rm avg}$
for some experimental scenarios considered in~\cite{Aliberti:2024fpq}, in~\tabref{tab:xs} we present the total cross sections with fiducial cuts, split into the various contributions defined in~\eqref{eq:Kfactor} and~\eqref{eq:defVPdeltas}.

\begin{table}[]
\centering
\begin{tabular}{
  l
  S[table-format=-2.6(1)]
  S[table-format=-2.6(1)]
  S[table-format=-2.6(1)]
}
\toprule
& {\makecell{KLOE-like \\ large-angle}} 
& {BESIII-like}
& {$B$} \\
\midrule
  $\sigma_0$~/~$\mu b$
    & 0.8792 & 0.26802 & 0.017498 \\[5pt]
  $\sigma_1$~/~$\mu b$
    & 1.0439 & 0.27486 & 0.016791 \\[5pt]
  $\sigma_1+\sigma^{(2)}_{\rm ISC}$~/~$\mu b$
    & 1.0423 & 0.27442 & 0.016693 \\[8pt]
\midrule
  $\delta K^{(1)}~/~\%$
    & 18.7 & 2.55 & -4.04 \\[3pt]
  $\delta K^{(1)}_{\rm ISC}~/~\%$
    & 7.13 & 2.91 & -1.63 \\[8pt]
  $\delta K^{(2)}_{\rm ISC}~/~\%$
    & -0.157 & -0.160 & -0.582 \\[3pt]
\midrule
  $\delta^{(2)}_{\rm VPC}[{\rm fac}]~/~\%$
    & -47 & -6.3 & 82 \\[3pt]
  $\delta^{(2)}_{\rm VPC}[\text{non-fac}]~/~\%$
    & 9.3 & 52 & 50 \\[3pt]
\bottomrule
  \hline
\end{tabular}
    \caption{
    Total cross section and relative contributions of the individual components for the KLOE-like large-angle, the BESIII-like, and the $B$ scenario with their fiducial cuts.
    The Monte Carlo error for the $\sigma_i$ is at most $\pm\,1$ in the last digit.
    }
    \label{tab:xs}
\end{table}

The \ac{NNLO} \ac{ISC} on the total cross section typically amount to a few permille w.r.t.~the full \ac{NLO} results.
The impact of both types of \ac{VPC} can be sizeable, in particular for scenarios with higher $\sqrt{s}$, and also reaches up to a few permille. Thus, they must not be neglected.
In general, by comparing $\delta K^{(1)}$ with $\delta K_{\rm ISC}^{(1)}$, we show that \ac{ISC} effects alone are not a good approximation of the total amount of corrections, except in certain regions of the phase space. 
Thus, a full \ac{NNLO} calculation is needed and, in fact, work is ongoing within the \mcmule{} team.

\subsection{The KLOE-like large-angle scenario} \label{sec:kloe-la-sc}

In this scenario we have $\sqrt{s}=1.02\,\GeV$ and request an energetic photon in the detector. The detailed acceptance cuts are 
\begin{subequations}\label{mcc:cutKLOE-la}
\begin{align}
&E_\gamma > 20\,\MeV& &\mbox{and}& 
&50^\circ\le\theta_\gamma\le 130^\circ\, , & \label{mcc:cutKLOE-la-y} 
\\
&|\vp^z_\pm| > 90\,\MeV \  \mbox{or}\  \vp^\perp_\pm > 160\,\MeV&
&\mbox{and}& &50^\circ\le\theta^\pm\le 130^\circ\, , & 
\\
&0.1\,\GeV^2 \le M_{\mu\mu}^2 \le 0.85\,\GeV^2\, , & 
\end{align}
\end{subequations}
where~\eqref{mcc:cutKLOE-la-y} implies the presence of at least one such photon. 
If further the cut 
\begin{align}
    \label{eq:cutmtrk}
    80\,\MeV \leq M_{\text{trk}} \leq 115\,\MeV
\end{align}
is applied to the distribution of $M_{\mu \mu}$, we obtain the observable $M_{\mu \mu}^c$.

\begin{figure}
    \centering
    \includegraphics[width=\textwidth]{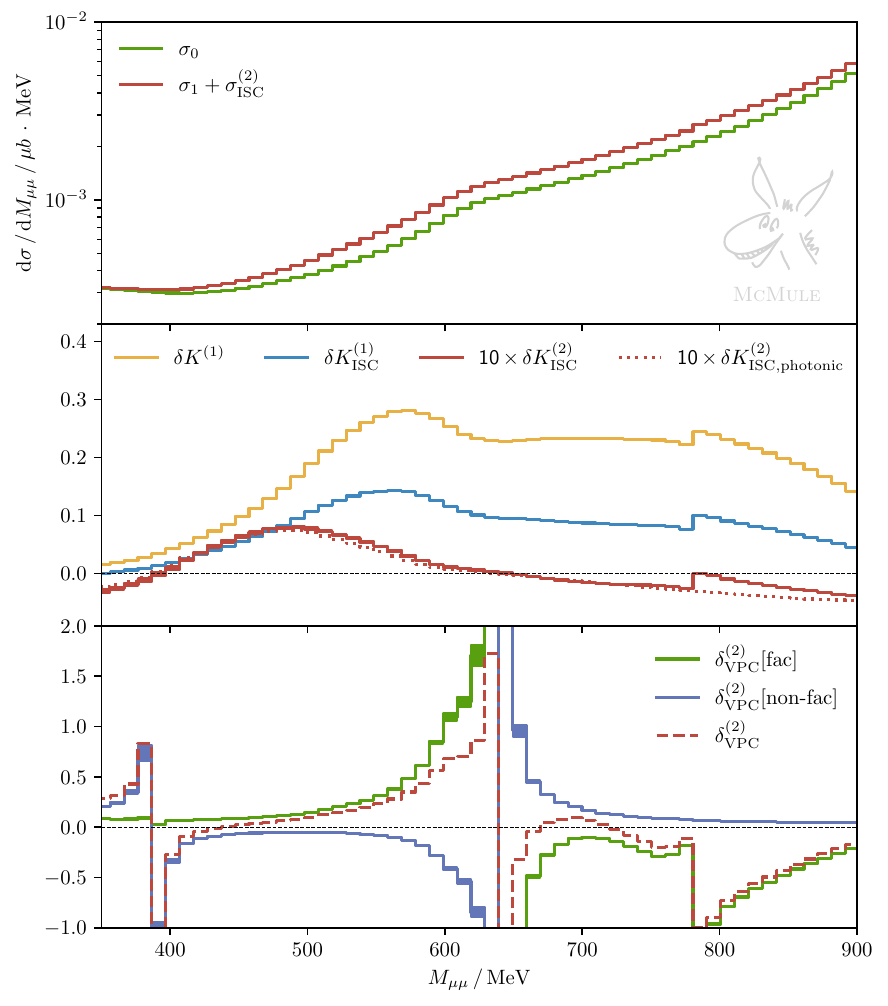}
    \caption{Results for the KLOE-like large-angle scenario for $M_{\mu\mu}$. The upper panel shows the \ac{LO} and \ac{NLO}+\ac{NNLO} \ac{ISC} distributions. The middle panel shows the various $\delta K$ factors as defined in~\eqref{eq:Kfactor}. The lower panel displays the relative contributions of factorisable and non-factorisable \ac{VPC} with respect to  photonic corrections at \ac{NNLO}.
 }
    \label{fig:kloe-mxx}
\end{figure}

\begin{figure}
    \centering
    \includegraphics[width=\textwidth]{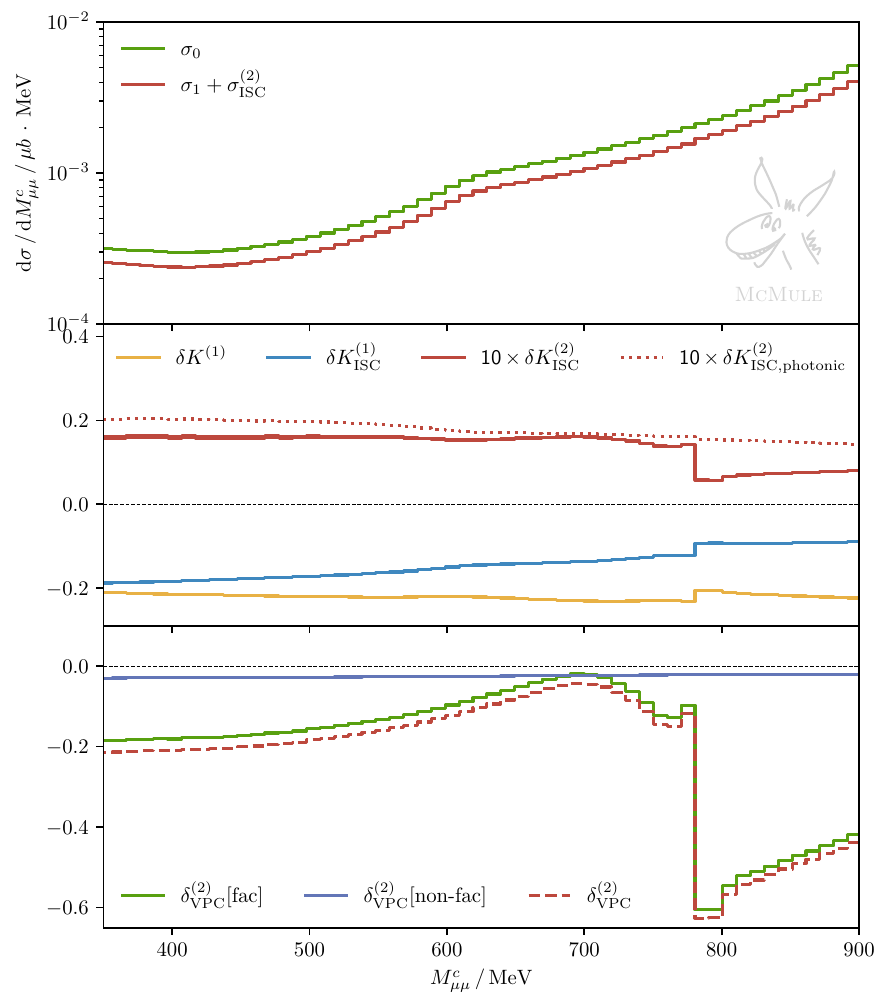}
    \caption{Results for the KLOE-like large-angle scenario as in Figure~\ref{fig:kloe-mxx} but with the additional track-mass cut~\eqref{eq:cutmtrk} applied.
}
    \label{fig:kloe-mxxc}
\end{figure}

\begin{figure}
    \centering
    \includegraphics[width=\textwidth]{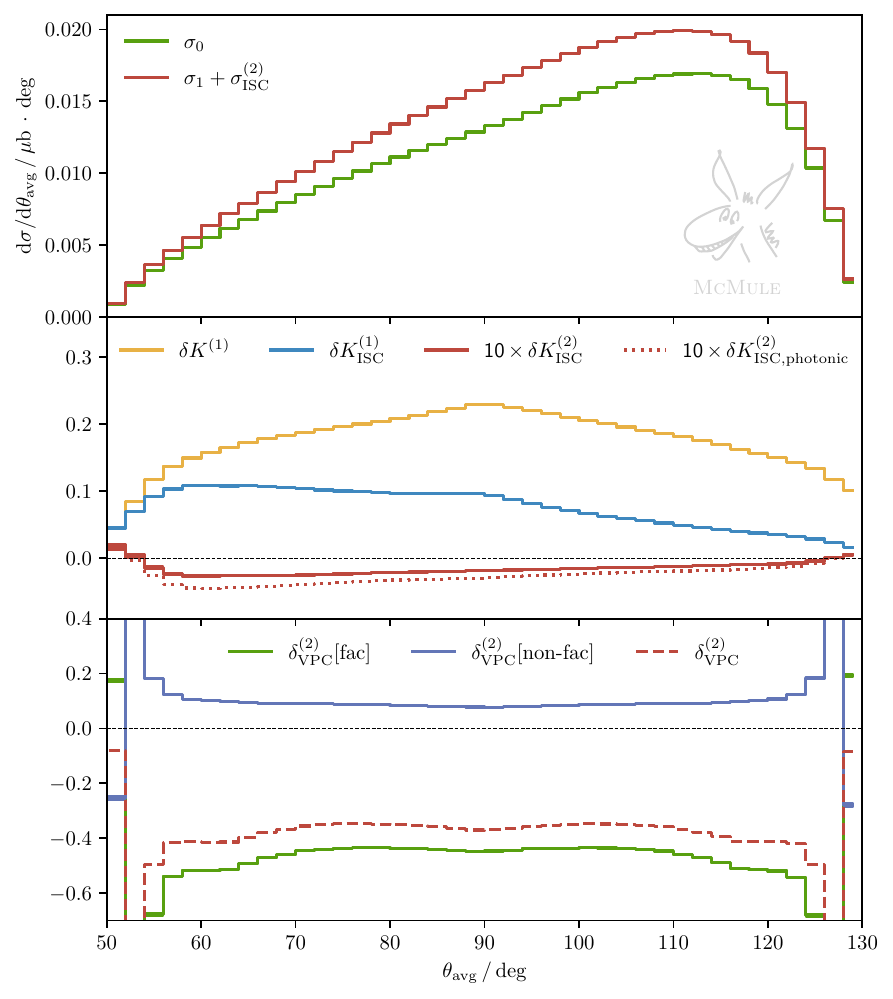}
    \caption{Results for the KLOE-like large-angle scenario for $\theta_{\text{avg}}$. Same distributions as in Figure~\ref{fig:kloe-mxx}.
 }
    \label{fig:kloe-thav}
\end{figure}

In Figures~\ref{fig:kloe-mxx}--\ref{fig:kloe-thav}, we present distributions for the KLOE-like large-angle scenario for $M_{\mu\mu}$ (Figure~\ref{fig:kloe-mxx}), $M_{\mu\mu}^c$ (Figure~\ref{fig:kloe-mxxc}), and $\theta_{\text{avg}}$ (Figure~\ref{fig:kloe-thav}). The same quantities are considered in all figures. The upper panel displays the \ac{LO} and the \ac{NLO}+\ac{NNLO} \ac{ISC} distributions. In the middle panel, we give the corresponding $\delta K$ factors to highlight the impact of the full corrections relative to the \ac{ISC} approximation. We also illustrate the impact of \ac{VPC} at \ac{NNLO} by comparing the full corrections to the photonic ones. The lower panel shows the ratio of \ac{NNLO} \ac{VPC} to \ac{NNLO} photonic corrections, which allows us to assess their relative importance.

For the $M_{\mu\mu}$ distribution shown in Figure~\ref{fig:kloe-mxx}, we observe a significant difference between the \ac{NLO} \ac{ISC} distribution and the full \ac{NLO} result.
The corrections of the latter (former) are of the order of $20\%$ ($10\%$) compared to the full \ac{LO}, respectively.
The difference between full and photonic \ac{NNLO} corrections is relatively small for $M_{\mu\mu}<770\,\MeV$. Around the $\rho$ resonance, however, the two results start to differ more significantly, reflecting the impact of the hadronic contribution to the \ac{VPC}. In the lower panel, we find that the dominant contribution from the \ac{VPC} above the $\rho$ resonance arises from factorisable diagrams.
All three curves display a peak between $600\,\MeV$ and $700\,\MeV$, which is a consequence of the \ac{NNLO} photonic corrections crossing zero in this region, as can be seen in the middle panel.

The situation is somewhat different in Figure~\ref{fig:kloe-mxxc}, where we additionally impose the cut~\eqref{eq:cutmtrk} on the track mass. For small values of $M_{\mu\mu}^c$, the difference between the \ac{NLO} \ac{ISC} and full \ac{NLO} results is considerably smaller. This region corresponds to the emission of a hard photon, with the selection cuts favouring hard-collinear photon emissions. At large values of the invariant mass, however, the \ac{ISC} approximation is no longer a good description of the full result. 
As before, the difference between full and photonic \ac{NNLO} corrections becomes sizeable above the $\rho$ resonance, indicating that \ac{VPC} cannot be ignored. Overall, the \ac{VPC} are almost entirely dominated by factorisable contributions, while the non-factorisable contributions account for only a few percent of the \ac{NNLO} correction.

Similar features can be observed in Figure~\ref{fig:kloe-thav}, where the \ac{NLO} \ac{ISC} approximation again deviates significantly from the full corrections. As before, the \ac{VPC} are largely accounted for by the factorisable contributions, which amount to about $40\%$ of the photonic \ac{NNLO} corrections.

\subsection{The BESIII-like scenario} \label{sec:bes-sc}

The centre-of-mass energy of the BESII-like scenario is $\sqrt{s}=4\,\GeV$. We request the detection of both muons and a photon, passing the selection cuts 
\begin{subequations}
\begin{align}\label{mcc:bescut}
&|\cos \theta^{\pm}|<0.93& &\mbox{and}&  &\vp_{\pm}^\perp >300\,\MeV\, ,& \\
& \big(|\cos\theta_\gamma| < 0.8 \ \mbox{and}\ E_\gamma > 25\,\MeV\big)&  &\mbox{or}& 
&\big(0.86 < |\cos \theta_\gamma| < 0.92\ \mbox{and} \ E_\gamma > 50\,\MeV\big) \, .
\end{align}
\end{subequations}
In addition, we require that precisely one detected photon has $E_\gamma \ge 400\,\MeV$. Thus, the emission of more than one (hard) photon is suppressed.

\begin{figure}
    \centering
    \includegraphics[width=\textwidth]{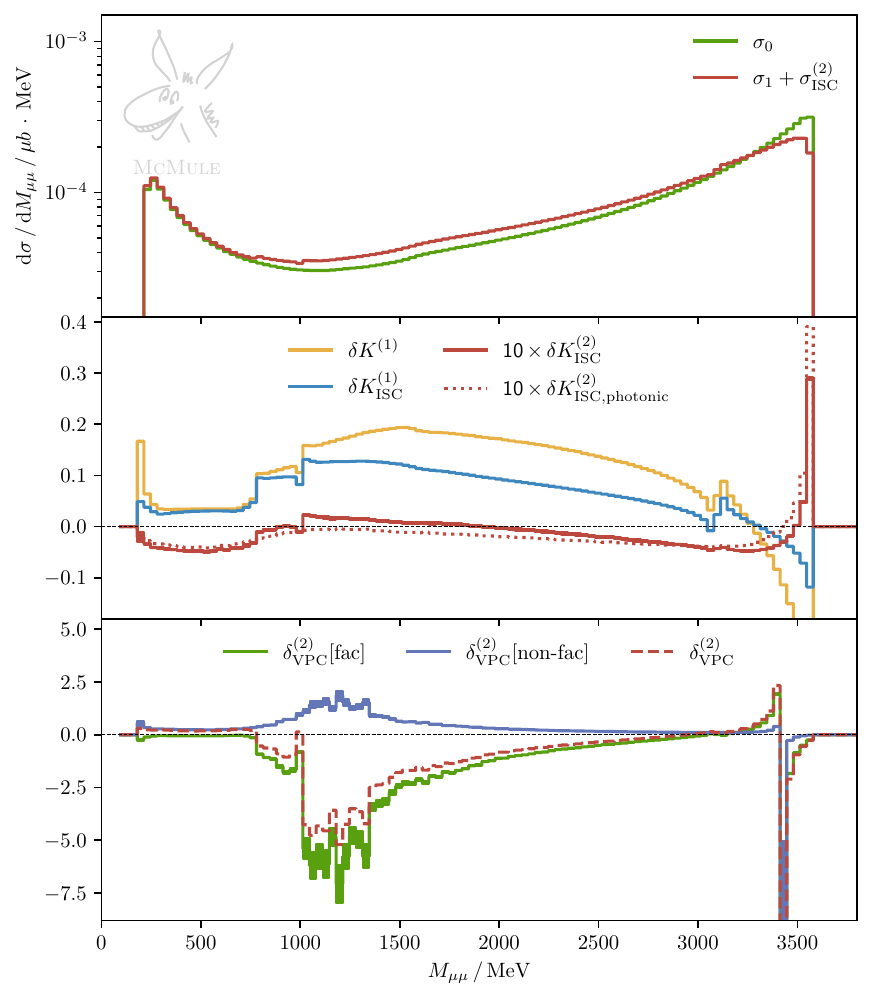}
    \caption{Results for the BESIII-like scenario for $M_{\mu\mu}$. Same distributions as in Figure~\ref{fig:kloe-mxx}.}
    \label{fig:bes-mxx}
\end{figure}

\begin{figure}
    \centering
    \includegraphics[width=\textwidth]{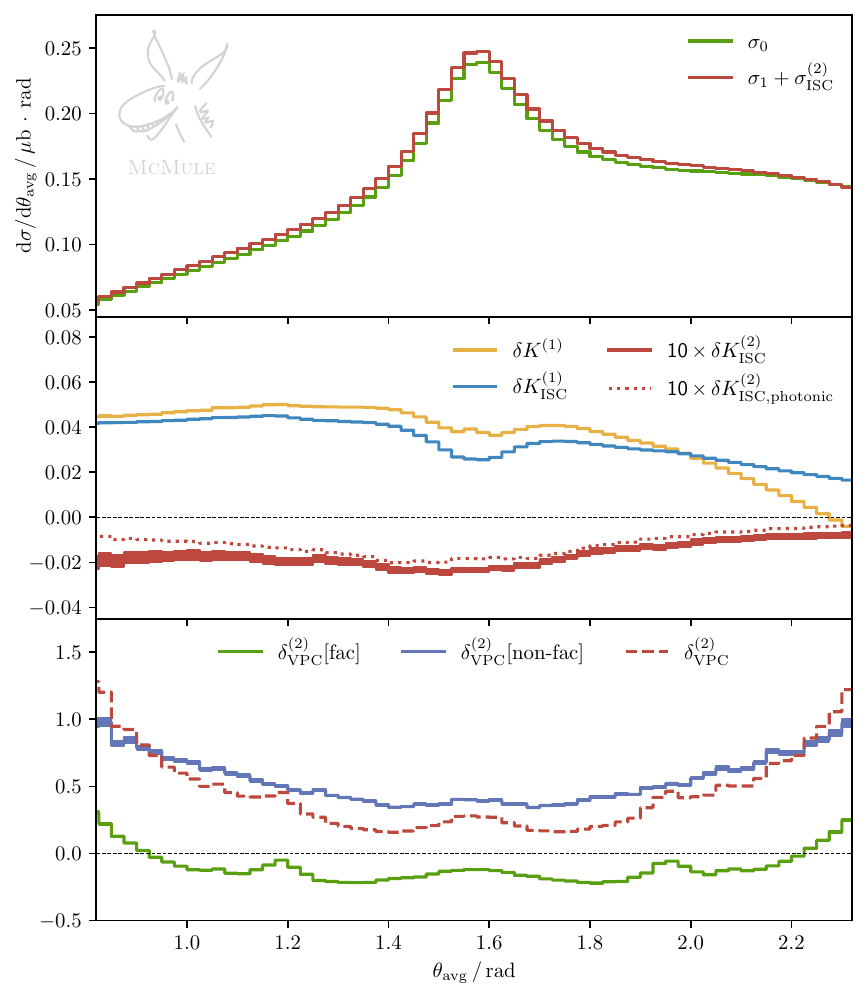}
    \caption{Results for the BESIII-like scenario for $\theta_{\text{avg}}$. Same distributions as in Figure~\ref{fig:kloe-mxx}.}
    \label{fig:bes-thav}
\end{figure}

In Figures~\ref{fig:bes-mxx} and~\ref{fig:bes-thav}, we show the distributions for the BESIII-like scenario for $M_{\mu\mu}$ and $\theta_{\text{avg}}$, respectively. In Figure~\ref{fig:bes-mxx}, we observe a similar behaviour to the KLOE-like large-angle scenario, with large \ac{NLO} corrections ranging from $10\%$ to $20\%$. For small $M_{\mu\mu}$, the \ac{NLO} \ac{ISC} and full \ac{NLO} results are in close agreement, while they show a clear deviation at large invariant masses. Moreover, we observe the $\rho$ resonance at $770\;\mathrm{MeV}$, as already seen in the KLOE-like large-angle scenario, together with additional hadronic resonances at higher invariant masses, namely the $\phi$ resonance at $\simeq\,1.0\,\GeV$ and the $J/\psi$ resonance at $\simeq\,3.1\,\GeV$. The \ac{NNLO} corrections itself are at the permille level.

While the \ac{VPC} are largely dominated by factorisable diagrams for a broad range of $M_{\mu\mu}$ values, in the region where the cross section is large, the non-factorisable \ac{VPC} are more important. Hence, we observe in the $\theta_{\text{avg}}$ distribution of Figure~\ref{fig:bes-thav} that the dominant contribution comes from non-factorisable diagrams. Overall, the corrections are much smaller in the $\theta_{\text{avg}}$ distribution, being of the order of a few percent at \ac{NLO} and $1$--$2$ permille at \ac{NNLO}. The middle panel further illustrates the importance of the \ac{VPC}, which are comparable in size to the photonic corrections.

\subsection{\texorpdfstring{\boldmath The $B$ scenario}{The B scenario}} \label{sec:b-sc}

Our final scenario is inspired by $B$ factories running at $\sqrt{s}=10\,\GeV$ with symmetric beams. For the photons and muons to be detected, we require
\begin{subequations}
\begin{align}\label{eq:Bcut}
&0.65\,\rad\le\theta^\pm\le 2.75\,\rad&  &\mbox{and}& &\vp_\pm > 1\,\GeV\, ,& \\
&0.6\,\rad\le\theta_\gamma\le 2.7\,\rad&  &\mbox{and}& &E_\gamma>3\,\GeV\, ,& \label{eq:Bcuty} \\
&\theta_{\gamma^{(h)},\widetilde\gamma} = 
\sphericalangle(\vec{p}_{\gamma^{(h)}},\vec{p}_{\widetilde\gamma}) < 0.3\,\rad&  
&\mbox{and}& &M_{\mu\mu\gamma} > 8\,\GeV\, .& \label{eq:Bcuth}
\end{align}
The second cut in~\eqref{eq:Bcuth} involves the hardest photon in the detector and suppresses additional (hard) photon emission.
\end{subequations}

\begin{figure}
    \centering
    \includegraphics[width=\textwidth]{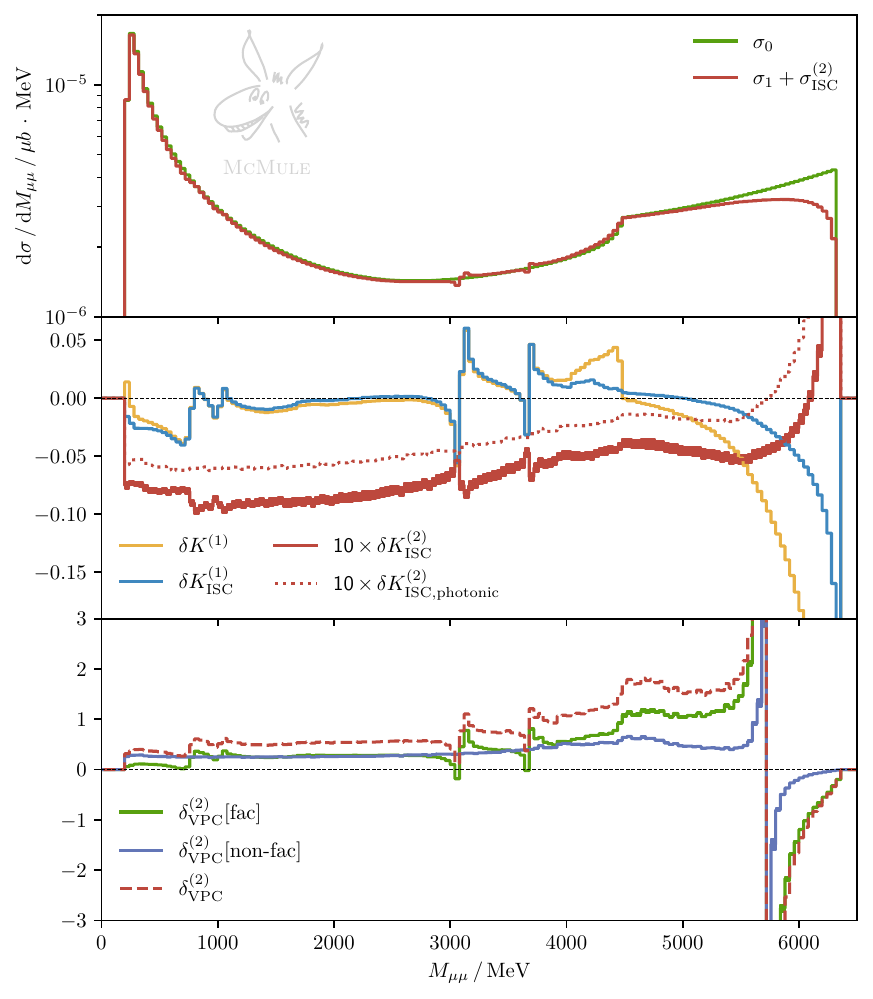}
    \caption{Results for the B-like scenario for $M_{\mu\mu}$. Same distributions as in Figure~\ref{fig:kloe-mxx}.}
    \label{fig:b-mxx}
\end{figure}

\begin{figure}
    \centering
    \includegraphics[width=\textwidth]{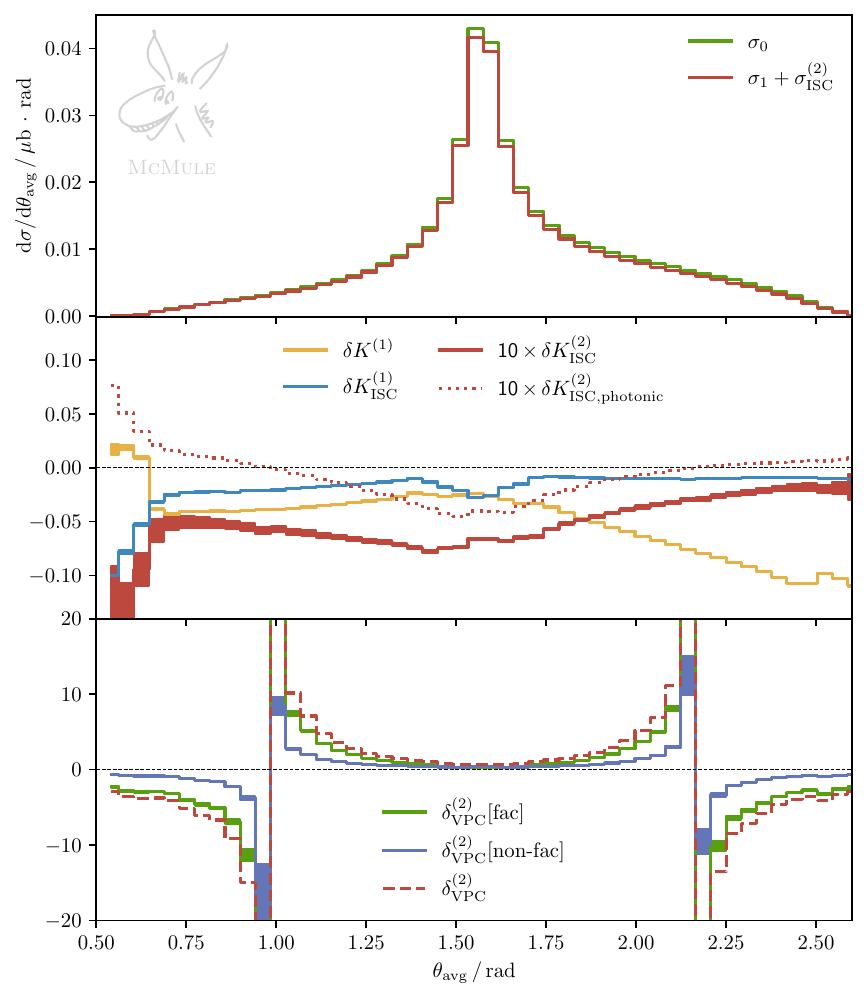}
    \caption{Results for the B-like scenario as a function of $\theta_{\text{avg}}$. Same distributions as in Figure~\ref{fig:kloe-mxx}.}
    \label{fig:b-thav}
\end{figure}

In Figures~\ref{fig:b-mxx} and~\ref{fig:b-thav}, we show the distributions for the $B$ scenario for $M_{\mu\mu}$ and $\theta_{\text{avg}}$, respectively. In Figure~\ref{fig:b-mxx}, the impact of the hadronic resonances is even more pronounced compared to the BESIII scenario. 
The additional $\psi(2S)$ resonance is reflected by a peak at $\simeq\,3.7\,\GeV$.
Both the factorisable and non-factorisable \ac{VPC} are comparable in size to the photonic corrections. The middle panel further illustrates the relevance of the \ac{VPC}, which account for $30$--$50\%$ of the \ac{NNLO} correction, depending on the value of the invariant mass. 
The \ac{ISC} is a reasonable approximation at \ac{NLO} for intermediate values of $M_{\mu\mu}$ but not for the extreme values for which the cross section is largest.
The feature of $\delta K^{(1)}$ at around $4.5\,\GeV$ is related to the cut on $\vp_\pm$.

Finally, Figure~\ref{fig:b-thav} confirms these observations: the \ac{NLO} \ac{ISC} approximation does not provide a good description of the full result, and the photonic corrections are not sufficient to describe the \ac{NNLO} result, with both the factorisable and non-factorisable \ac{VPC} providing sizeable contributions.

\section{Outlook} \label{sec:out}

The core part of the results presented in this article is the subprocess $e^+\,\,e^-\to\gamma\,\gamma^*$ at \ac{NNLO} in QED, including \ac{VPC}. We have used this to present \ac{ISC} to the $2\to{3}$ process with a muon pair in the final state. The only restriction to the applicability is related to the use of massification for the $m_e$ dependence of the photonic two-loop amplitude. This requires $m_e$ to be much smaller than any other scale of the process. Scenarios where a (hard) photon is emitted collinear to the electron or positron violate this condition. A computation of the two-loop amplitude $e^+\,\,e^-\to\gamma\,\gamma^*$  with full $m_e$ dependence would be very useful to avoid this restriction. Such a calculation could also be used to extend the collinear limit of massive one-loop amplitudes~\cite{Engel:2021ccn} to the two-loop case.

The \ac{NNLO} building block  $e^+\,\,e^-\to\gamma\,\gamma^*$ has several further applications that will be considered in the future. 
First, it can be combined with the recent \ac{NLO} calculation~\cite{Fang:2026naz} of $e^+\,e^-\to\pi^+\,\pi^-\,\gamma$. 
While for $e^+\,e^-\to\mu^+\,\mu^-\,\gamma$ the \ac{ISC} approximation is often not well justified, in the case of pion final states the \ac{ISC} are typically the by far dominant contribution~\cite{Fang:2026naz}. 
Another application is to consider the radiative tail in lepton-proton scattering $e^-\,p\to{e}^-\,p\,\gamma$, e.g.~\cite{A1:2015wgf,A1:2019mrv}.

In a broader context, these results are part of the full \ac{NNLO} $e^+\,e^-\to\mu^+\,\mu^-\,\gamma$ calculation that is required for a precision at the permille level. They can also be seen as part of a \acs{NNNLO} calculation for $e^+\,e^-\to\gamma^*$. In this case, an integration over the full photon phase space is required. Again, this includes a region where massification is questionable. Since also the crossed version $e\to{e}\,\gamma^*$ of the \acs{NNNLO} building block will have several phenomenologically relevant applications, a careful analysis of the massification error will be important, as long as the full result is not available. In this context, we mention first steps towards a \acs{NNNLO} calculation of muon-electron~\cite{LoI,Banerjee:2020tdt} or lepton-proton scattering~\cite{Afanasev:2023gev}.

With these results we have made another step in improving fixed-order calculations for low-energy cross sections
A first comparison of \ac{ISC} beyond \ac{NLO} (without \ac{VPC}) shows that the total prediction of \mcmule{}'s \ac{NNLO} can differ w.r.t.~\textsc{BabaYaga@NLO}'s \ac{NLO}$+$\ac{PS}~\cite{Budassi:2026lmr} at $\mathcal{O}(0.1\,\%)$ for some regions of the phase space~\cite{Kollatzsch:2026TI,Ghilardi:2026Torino}.
A dedicated comparison is planned for the future within the RadioMonteCarLow~2 effort.

\subsection*{Acknowledgement}
It is a pleasure to thank the \textsc{BabaYaga@NLO} team, especially Carlo~Carloni~Calame, Marco~Ghilardi and Fulvio~Piccinini for providing their \ac{NLO}+\ac{PS} prediction~\cite{Budassi:2026lmr} restricted to \ac{ISC}.

YF (10006722), SK (207386) and SG (207386 and 10006722) acknowledge support by the Swiss National Science Foundation (SNSF) under the respective grants. MR acknowledges support by the Italian Ministry of Universities and Research (MUR) through grants PRIN 2022BCXSW9 and FIS (CUP: D53C24005480001, FLAME).

\bibliographystyle{JHEP}
\bibliography{mmy}

\end{document}